\documentclass[prc,aps,preprint,showpacs]{revtex4}
\usepackage{graphicx}
\newcommand{\simgeq}{\; \raisebox{-0.4ex}{\tiny$\stackrel
{{\textstyle>}}{\sim}$}\;}
\newcommand{\simleq}{\; \raisebox{-0.4ex}{\tiny$\stackrel
{{\textstyle<}}{\sim}$}\;}
\begin{document}
\title{Interpretation of Coulomb breakup of $^{31}$Ne in terms of deformation} 

\author{ Ikuko Hamamoto$^{1,2}$ }

\affiliation{
$^{1}$ {\it Division of Mathematical Physics, Lund Institute of Technology 
at the University of Lund, Lund, Sweden}   \\
$^{2}$ {\it The Niels Bohr Institute, Blegdamsvej 17, 
Copenhagen \O,
DK-2100, Denmark} \\ 
}




\begin{abstract}
The recent experimental data on Coulomb breakup of the nucleus 
$^{31}$Ne are interpreted in terms of deformation.  The measured large
one-neutron removal cross-section 
indicates that the ground state of $^{31}$Ne is either s- or
p-halo.  The data can be most easily interpreted as the spin of the ground 
state being 3/2$^-$ coming from either the Nilsson level [330 1/2] or 
[321 3/2] depending on the neutron separation energy $S_n$.  However, the
possibility of 
1/2$^{+}$ coming from [200 1/2] is not excluded.   It is suggested that if the 
large ambiguity 
in the measured value of $S_n$ of $^{31}$Ne, 0.29$\pm1.64$ MeV, 
can be reduced 
by an order of magnitude, say to be $\pm$100 keV, 
one may get a clear picture of the
spin-parity of the halo ground state.   
\end{abstract}


\maketitle

Recent experimental data obtained by using radioactive ion beams 
reveal that the neutron numbers such as N=8, 20 and 28 are no longer magic
numbers in some nuclei towards the neutron drip line.  
Those neutron-rich nuclei 
can well be interpreted as being deformed and are often called nuclei 
in the island of inversion. 
One-neutron removal cross sections of 
a very neutron-rich nucleus 
$^{31}$Ne due to the Coulomb breakup are reported in \cite{TN09} and found 
to provide an evidence of the soft E1 excitation. 
The analysis of the cross section in Ref. \cite{TN09} is based on 
one-particle wave-functions in the spherical Woods-Saxon potential.
It was concluded that the measured large one-neutron removal cross sections 
were consistent with a $p_{3/2}$ or $s_{1/2}$ neutron halo when a
spectroscopic factor considerably smaller than unity was introduced. 
The heaviest halo nucleus so far experimentally established is $^{19}$C 
with the
ground-state spin of 1/2$^{+}$, namely an s-halo nucleus.  
While the nucleus $^{19}$C may be interpreted as a spherical halo nucleus, 
the data on $^{31}$Ne in \cite{TN09} may be most easily 
interpreted in terms of the deformation or the one-particle motion 
in the deformed mean-field, since
the spin-parity of the halo ground-state cannot be 7/2$^-$ which is expected 
for the 21st neutron for spherical shape. 
We note that the analysis of the spectroscopic data on light 
mirror nuclei, 
$^{25}_{12}$Mg$_{13}$ and $^{25}_{13}$Al$_{12}$, is successfully performed 
in \cite{BM75} in terms of the deformed mean field using 
Nilsson orbits occupied by the 13th nucleon.
In the present Communication we attempt the interpretation of the data in
\cite{TN09} based on a deformed
mean-field, using one-particle neutron wave-functions  
obtained by taking properly into account the weak binding.  

The measured low excitation energies of the first 2$^{+}$ state of both 
$^{30}$Ne \cite{YY03} and $^{32}$Ne \cite{PD09} are consistent with the picture 
that the Ne isotope with these neutron-numbers lies inside the island of 
inversion.  
Moreover, the large Coulomb breakup cross section reported in \cite{TN09}, which
clearly indicates the halo nature of the ground state of $^{31}$Ne, suggests 
the efficient contribution by $s$ or $p$ component of the 21st neutron in the
deformed mean field.  These
observations invite us to carry out the analysis of the data on 
$^{31}$Ne in terms of deformation.  
It should be also noted that the measured magnetic moment of the ground state of
$^{33}_{12}$Mg$_{21}$, which has the same neutron-number N=21, 
is reported in Ref.
\cite{DY07} and is consistent with the interpretation of $I^{\pi}$=3/2$^-$, 
while both neighboring even-even nuclei, $^{32}$Mg and $^{34}$Mg, are well
interpreted as being deformed.  However, since the neutron separation energy of
$^{33}_{12}$Mg$_{21}$ is 2.22 MeV, a neutron halo is not expected. 

Low-lying states of odd-A medium-heavy deformed nuclei are in a good 
approximation expressed by one
(quasi)particle moving in the deformed field produced 
by the even-even
core \cite{BM75}.  One-particle picture works much better in deformed
nuclei than in spherical nuclei, because the major part of the long-range 
residual
interaction in the spherical mean field can be included in the deformed
mean field.
In contrast, when the spherical shell model is applied to nuclei 
in the island of inversion, the resultant wave-functions are not easy to be
predicted due to the complicated configuration mixing.  
Furthermore, since harmonic-oscillator wave-functions are always
used in the traditional shell model calculation, the applicability of the
calculations to the description of halo nuclei can be questioned. 

We start with our formulation for spherical shape. 
The neutron bound-state wave-function is an eigenfunction of the Woods-Saxon
potential with an energy eigenvalue $\varepsilon < 0$
\begin{equation}
\mid \Phi_{\varepsilon}^{(b)}: \ell j m \rangle = 
\frac{1}{r} R_{n \ell j}^{(b)}(\varepsilon,r) 
[Y_{\ell} \otimes \chi_{1/2}]_{jm} 
\end{equation}
where
\begin{equation}
R_{n \ell j}^{(b)}(\varepsilon,r) \propto \alpha r \, h_{\ell}(\alpha r), 
\end{equation}
with 
\begin{eqnarray}
h_{\ell}(-iz) & \equiv & j_{\ell}(z) \, + \, i \, n_{\ell}(z) \\
\alpha^2 & = & - \, \frac{2 \mu \varepsilon}{\hbar^2} \\
\int_{0}^{\infty} dr \mid R_{n \ell j}^{(b)} (\varepsilon, r) \mid^2 & = & 1
\end{eqnarray}
We choose to express one-particle neutron wave-functions 
in the continuum using real energy variable $\varepsilon > 0$  
\begin{eqnarray}
\mid \Phi_{\varepsilon} ^{(c)}: \ell j m \rangle & = &
\frac{1}{r} R_{\ell j}^{(c)}(\varepsilon, r) [Y_{\ell} \otimes 
\chi_{1/2}]_{jm} \\ 
& = & 
\sqrt{\frac{2 \mu}{\hbar^2 \pi k}} \, 
(cos(\delta_{\ell j}) \, k \, j_{\ell}(kr)-sin(\delta_{\ell j}) \, k \, 
n_{\ell}(kr)) \, 
[Y_{\ell} \otimes \chi_{1/2}]_{jm}
\end{eqnarray}
where $\mu$ expresses the reduced mass and 
\begin{equation}
k^2 = \frac{2 \mu \varepsilon}{\hbar^2}
\end{equation}
while the normalization is expressed as 
\begin{equation}
\int_{0}^{\infty}dr R_{\ell j}^{(c)}(\varepsilon, r)R_{\ell j}^{(c)}
(\varepsilon',r)
=\delta(\varepsilon - \varepsilon') .
\end{equation}
The B(E1) value of the E1 excitation from a bound one-particle ($j_b$) level to
a continuum one-particle ($j_c$) level keeping the even-even core in the ground
state as a spectator is written as 
\begin{equation}
\frac{dB(E1, b(j_b) \rightarrow c(j_c))}{dE} = (Z_{eff} e)^2 \, 
\frac{2j_{c}+1}{4 \pi} \, [C(j_c, j_b, 1 ; \frac{1}{2}, \frac{-1}{2}, 0)]^2 \, 
\mid \int rdr 
R_{j_b}^{(b)}(\varepsilon_b, r) R_{j_c}^{(c)}(\varepsilon_c, r) \mid ^2
\label{eq:BE1}
\end{equation}

First, it is found that the Coulomb breakup by the soft E1 excitation 
of 
halo neutrons occurs far outside the nucleus $^{31}$Ne, noting that no
one-particle resonance with the relevant angular momentum is present in the 
low-energy continuum.  
In Fig. 1 we plot the 2$p_{3/2}$ radial wave-function with
the eigenenergy $\varepsilon_{b}$=$-$300 keV, the continuum $s_{1/2}$ radial
wave-function with $\varepsilon_{c}$=+60 keV, and the product of those two
wave-functions multiplied by radial variable. 
The value of $\varepsilon_{c}$ is chosen as an example,  because for
$\varepsilon_{b}$=$-$300 keV the quantity in Eq. (\ref{eq:BE1}) reaches the
maximum around $\varepsilon_{c}$=+60 keV \cite{NLV05}. 
From Fig. 1 it is seen that the major part of the soft E1 matrix-element 
comes from the region of
$r$=8-40 fm, far outside of the core nucleus $^{30}$Ne. 
A similar result is obtained also for soft 
E1 excitations of the halo $p$
neutron to continuum $d$ levels, though the Coulomb breakup cross-section is
much smaller than that of the excitation to $s$ levels.
Both the tail shape and the amplitude of the wave function of the halo neutron
are very important for the cross sections of Coulomb breakup, 
while the wave functions inside $^{31}$Ne may
hardly play a role. 

In the phenomena where the structure of deformed wave-functions plays a role, 
the expression in Eq. (\ref{eq:BE1}) is replaced by the reduced matrix-element
in Eq. (4-91) of Ref. \cite{BM75} together with Eq. (1A-67) of Ref. \cite{BM69}. 
Continuum one-particle wave-functions in a deformed mean-field are calculated
in terms of eigenphase \cite{IH05,IH06}.  For given $\varepsilon_{c}$ and
$\Omega^{\pi}$ values there are a number of independent 
one-particle wave-functions, the
number of which is equal to that of eigenphases, and all possible contributions
must be calculated and summed up. 
However, when soft E1 excitations should be estimated which produce 
such large Coulomb breakup cross-sections as those in
\cite{TN09}, we may conveniently 
use the halo one-neutron wave-function taken from 
the $s$ or $p$ component of deformed Nilsson levels (an approximation in terms
of ''spectroscopic factor''),
while one-particle wave-functions in the continuum are estimated for the
spherical part of the Woods-Saxon potential.

In Fig. 2 the Nilsson diagram is shown, of which the parameters are
approximately adjusted to 
the ($^{30}$Ne+n) system.  The calculation was done in the same manner as in
\cite{IH07}.   At $\beta$=0 the 1$f_{7/2}$ one-neutron resonance
is found at 2.40 MeV with the width of 0.224 MeV, while 
neither the 2$p_{3/2}$ nor 2$p_{1/2}$ resonance defined by 
the eigenphase formalism \cite{IH05,IH06} is obtained.
However, the complicated non-linear behavior of the [330 1/2] resonant 
level for 
$0.1 < \beta < 0.2$ in the continuum (denoted by a dotted curve) 
indicates that the resonant-like 
component with $\Omega^{\pi}$=1/2$^{-}$ coming from 2$p_{3/2}$ is present 
around the energy region.  Indeed the
2$p_{3/2}$ resonance lying lower than the 1$f_{7/2}$ one is found 
if we use a slightly more attractive Woods-Saxon potential.  
For the parameters used in Fig.2 the Nilsson level which is to be occupied by
the 21st neutron is [330 1/2] for $0.22 \simleq \beta \simleq 0.30$, 
[202 3/2] for $0.30 \simleq \beta \simleq 0.40$, [321 3/2] for 
$0.40 \simleq \beta \simleq 0.59$ and [200 1/2] for $\beta \simgeq 0.59$.
Varying the parameters of the one-body potential within a reasonable range, any
Nilsson levels other than those four levels are hardly obtained for the 21st
neutron.
The spin-parity of the lowest state is I$^{\pi}$=3/2$^-$ for [330 1/2] due to
the decoupling parameter lying between $-2$ (for $p_{3/2}$) and $-4$ (for the
$f_{7/2}$), 
3/2$^+$ for [202 3/2], 
3/2$^-$ for [321 3/2] and 1/2$^+$ for [200 1/2].
Among those four Nilsson orbitals the [202 3/2] level is excluded
as a candidate for the configuration of the ground state of $^{31}$Ne, 
because the smallest orbital-angular-momentum in the wave function of [202 3/2]
is $\ell$=2, which makes very little halo.
Examining Fig. 2 we may also note 
that the presence of $^{31}$Ne inside the neutron 
drip line is possibly realized by the deformation which is created 
by Jahn-Teller effect 
due to the near degeneracy of 1$f_{7/2}$, 2$p_{3/2}$ and 2$p_{1/2}$ shells 
in the continuum for spherical shape.

In Figs. 3(a) and 3(b) the probabilities of appreciable components 
of the [330 1/2] and [321 3/2] levels calculated at
$\beta$=0.3 and 0.5, respectively, are shown, while the channels of  
$p_{1/2}$ (only in [330 1/2]), $p_{3/2}$, $f_{5/2}$, $f_{7/2}$, $h_{9/2}$ 
and $h_{11/2}$ 
are included in the calculation.
The radius of the Woods-Saxon potential is fixed, while 
the depth is adjusted so as to obtain respective Nilsson levels as 
eigenstates of the deformed potential. 
As shown in Refs. \cite{MNA97,IH04}, the $p$ components in $\Omega^{\pi}$ = 
1/2$^-$ and 3/2$^-$ Nilsson levels increase as the binding energies approach
zero, though the probabilities at zero energies depend on Nilsson levels. 
This is in contrast to the fact that the probability of  
the $s$ component in $\Omega^{\pi}$=1/2$^{+}$ Nilsson levels 
becomes always unity as the binding energy approaches zero. 
For example, at $\varepsilon_{\Omega}$=$-$300 keV the probability of 
the $p_{3/2}$ component 
in the [330 1/2] for $\beta$=0.3 and [321 3/2] for $\beta$=0.5 
levels is 0.5225 and 0.2534, respectively.
For reference, at $\varepsilon_{\Omega}$=$-$300 keV the probability of 
the $s_{1/2}$ component in
the [200 1/2] level for $\beta$=0.5 is 0.67.

Now, for example, the shape of the 
radial wave-function of the $p_{3/2}$ component of 
[330 1/2] at $\beta$=0.3 is not the same as that of the bound 2$p_{3/2}$
level, since the latter is an eigenstate
of a given spherical potential while the former is not.
Moreover, two different spherical potentials lead to different 
radial wave-functions of $s_{1/2}$ at a given energy in 
the continuum.  
These differences may induce a non-negligible change in 
the resulting B(E1) values, even after the normalization of the 2$p_{3/2}$
wave-function is adjusted to be the same as the probability of the $p_{3/2}$
component in [330 1/2]. 
In Fig. 4 we show the squared radial integral on the r.h.s. of 
Eq. (\ref{eq:BE1}) as a function of the continuum $s_{1/2}$ energy
$\varepsilon_c$, which is calculated using the following 
two kinds of $p_{3/2}$ bound-state wave-functions;  
(i) the 2$p_{3/2}$ wave function with the energy eigenvalue $-$300 keV 
for a spherical potential and a normalization of 0.5225; 
(ii) the $p_{3/2}$ component of [330 1/2] which has the energy of $-$300 keV in
the deformed potential with $\beta$=0.3. 
The appreciable difference in dB(E1)/dE values of the
two cases appears only for small $\varepsilon_c$ values, and the integration of
the two curves over $\varepsilon_c$ up till 2 MeV gives a difference of 
about 15 percent. 
Generally speaking, if we replace the $p_{3/2}$ component of the [330 1/2] or 
[321 3/2] level by 
the 2$p_{3/2}$ wave function 
with the same neutron-binding and normalization, the E1 transition 
2$p_{3/2}$ $\rightarrow$ $s_{1/2}$ has a larger B(E1)-value than that of 
$p_{3/2}$ $\rightarrow$ $s_{1/2}$.  
''Larger'' or ''smaller'' depends on whether the spherical (2$p_{3/2}$) 
eigenstate lies energetically higher or lower than the one-particle level 
in the Nilsson diagram for the relevant value of $\beta \neq 0$.
The larger the energy difference, the larger difference in B(E1) values
appears which comes from the different radial shape of the two wave-functions. 

Combining Figs. 3(a) and 3(b) with Fig. 2 of Ref. \cite{TN09}, 
for a given $S_n$ value we may find 
the relevant Nilsson level and the spin-parity of the ground state
of the halo nucleus $^{31}$Ne, which are consistent with 
the observed Coulomb breakup cross-section. 
After taking into account both 
possible ambiguities in the parameters of the one-body
potential and the approximation in terms of ''spectroscopic factor'' of $s$ or
$p$ neutrons, we may conclude at least the following;
The ground state has I$^{\pi}$=1/2$^+$ coming from the Nilsson level [200 1/2],
if $S_n$ is appreciably larger than 500 keV. In this case the relevant
deformation is expected to be very large, $\beta \simgeq 0.6$. 
If $S_n$ is smaller than 500 keV, the ground state can be $p$-wave halo and
I$^{\pi}$=3/2$^-$.  If $S_n$ is smaller than 200 keV, the relevant Nilsson level
may be [321 3/2].  Otherwise, it is [330 1/2].  

To improve the accuracy of the measured $S_n$ value much better 
than the available one \cite{BJ07}, 
$S_n$=0.29$\pm$1.64 MeV, can clarify 
the I$^{\pi}$ value when it is combined with the data in \cite{TN09}.  
On the other hand, for example, 
the measurement of the magnetic moment of $^{31}$Ne may not
clearly pin down the spin-parity, since the estimated magnetic moment coming
from the
N=21st neutron in a deformed core is anyway negative and lies in the range of 
$-0.4 \simleq \mu / \mu_{N} \simleq -1.0$ for 
possible states with $I^{\pi}$=3/2$^-$ coming from either [330 1/2] or 
[321 3/2] Nilsson orbits and those with $I^{\pi}$=1/2$^+$ from [200 1/2].

In the present analysis the possibility of 
excited states of $^{30}$Ne after the 
Coulomb breakup is not included, however, 
the related Coulomb breakup cross-section is relatively
small and lies within the ambiguities in the model calculation. 
The possible many-body pair-correlation is not included either. 
This is because, 
first of all, the halo neutron wave-function extends so much beyond the
core nucleus that it couples weakly to the pairing field provided by the
well-bound core nucleons.  Secondly, in the ground state of odd-N nuclei the
relevant neutron single-particle energy lies very close to the
Fermi level.  Then, irrespective of the nature of one-particle orbits, the
occupation probability of the doubly-degenerate neutron level obtained from
solving the HFB equation is approximately equal to 0.5 \cite{HM03,IH06a}. 
Therefore, the contribution of halo neutrons to the Coulomb breakup which is
estimated in the present article is expected to work as a first approximation. 

In conclusion, it is shown that the observed large Coulomb breakup cross section
of $^{31}$Ne in Ref. \cite{TN09} is interpreted most easily and simply 
in terms of
p-wave neutron-halo together with the deformed core $^{30}$Ne.
The measurement of $S_n$ with an accuracy much better than the presently 
available one will clarify the spin-parity of the ground state of $^{31}$Ne.

Fruitful discussions with Prof. T. Nakamura are much appreciated.

\newpage

\noindent
{\bf\large Figure captions}\\
\begin{description}
\item[{\rm Figure 1 :}]
The integrand of the radial integral on the r.h.s. of Eq. (\ref{eq:BE1}) as a
function of radial variable $r$. 
Namely, the product (solid curve) of radial variable, the bound 
2$p_{3/2}$ radial 
wave-function $R_{2p_{3/2}}^{(b)}(\varepsilon = -300 \,\mbox{keV}, r)$ 
(dashed curve) in
unit of fm$^{-1/2}$, 
and the continuum 
$s_{1/2}$ radial wave-function $R_{s_{1/2}}^{(c)}(\varepsilon = 
 +60 \, \mbox{keV}, r)$ (dotted curve) in unit of fm$^{-1/2}$ MeV$^{-1/2}$.  
The unit of the ordinate is for the product, while the figures written along the
ordinate can be used also for the radial wave-functions, $R^{(b)}_{2p_{3/2}}$ 
and $R^{(c)}_{s_{1/2}}$, in respective units.
The radial integration
of the solid curve gives the radial matrix-element of the E1 transition, 
2$p_{3/2}$ $\rightarrow$ $s_{1/2}$. 
The depth of the Woods-Saxon potential is $-$46.75 MeV for which the
eigenstate of the 2$p_{3/2}$ neutron is obtained 
at $-$300 keV, while the diffuseness and
the radius used are 0.67 fm and 3.946 fm (for A=30), respectively.
\end{description}

\begin{description}
\item[{\rm Figure 2 :}]
Neutron one-particle levels in Woods-Saxon potentials 
as a function of quadrupole
deformation parameter $\beta$.  The potential depth is approximately 
adjusted so that the 21st neutron of the prolately deformed nucleus $^{31}$Ne 
can be a halo neutron.  The depth, the diffuseness 
and the radius of 
the potential are $-$39 MeV, 0.67 fm and 3.946 fm (for A=30), respectively. 
Positive-parity levels are plotted by solid curves, while asymptotic quantum
numbers [N n$_z$ $\Lambda$ $\Omega$] are denoted for bound levels. 
See the text for details. 
\end{description}

\noindent
\begin{description}
\item[{\rm Figure 3 :}]
(a) Calculated probabilities of the major components of the [330 1/2] level 
with $\beta$=0.3 as a function of energy eigenvalue $\varepsilon_{\Omega}$; 
(b) Calculated probabilities of the major components of the [321 3/2] level 
with $\beta$=0.5 as a function of energy eigenvalue $\varepsilon_{\Omega}$. 
The potential depth is
adjusted to obtain respective $\varepsilon_{\Omega}$ values as energy
eigenvalues of
the deformed potential.  
The diffuseness and the radius of the potential are 0.67 fm and 3.946 fm (for
A=30).
\end{description}

\begin{description}
\item[{\rm Figure 4 :}]
Estimate of $ \mid \int \, r \, dr \, R_{p_{3/2}}^{(b)}(-300 
\mbox{keV}, r) \, R_{s_{1/2}}^{(c)}(\varepsilon_c, r) \mid^2 $ 
on the r.h.s. of Eq. (\ref{eq:BE1}) as a function of $\varepsilon_c$.
The solid curve is obtained by using the $p_{3/2}$ wave-function 
taken from the $p_{3/2}$ 
component of the [330 1/2] level, 
which is bound at $-$300 keV for $\beta$=0.3.  
The probability of the $p_{3/2}$ component is 0.5225. 
The dotted curve is calculated
using 
the 2$p_{3/2}$ wave-function with the energy eigenvalue 
$-$300 keV, which is normalized to 0.5225.  The depth of the Woods-Saxon
potential of the former is $-$38.39 MeV, while that of the latter is $-$46.75
MeV.  
\end{description}

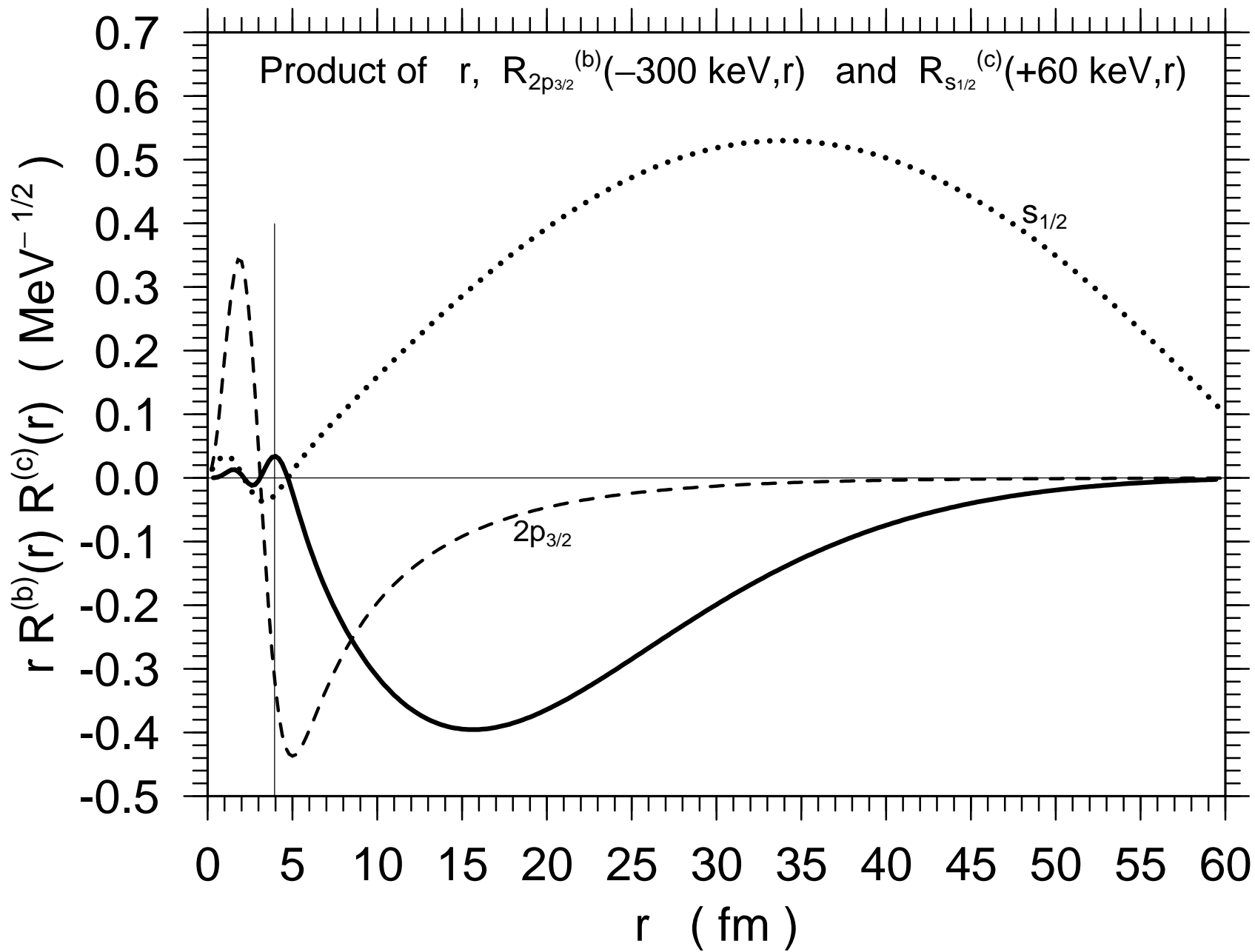
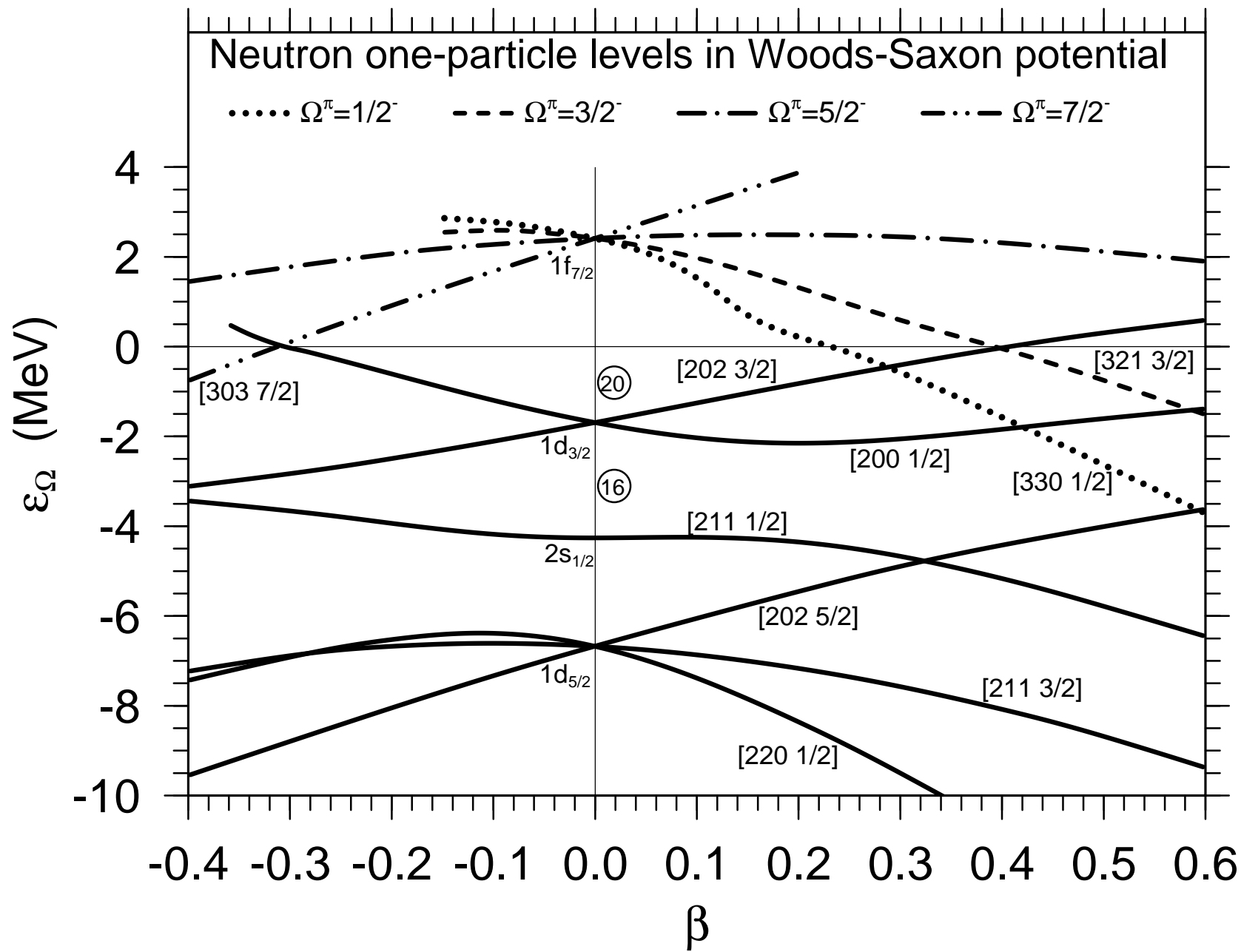
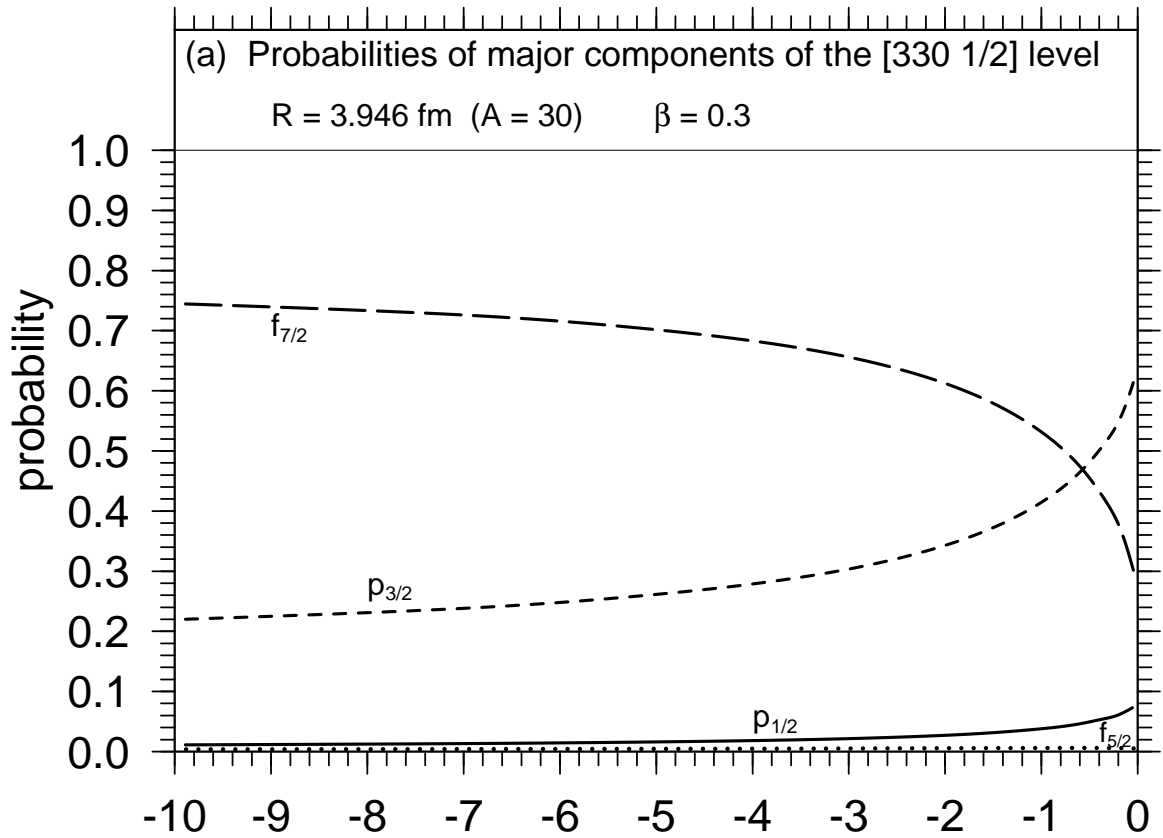
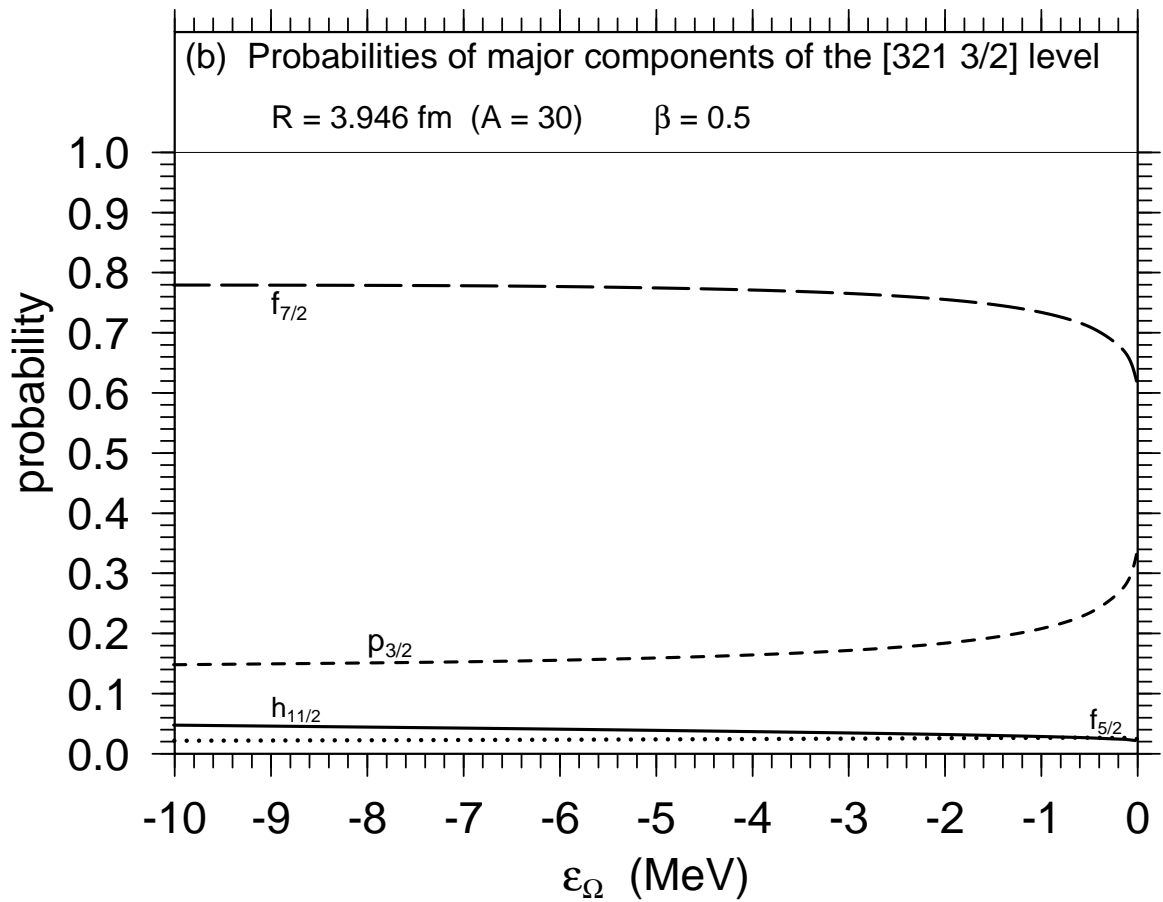
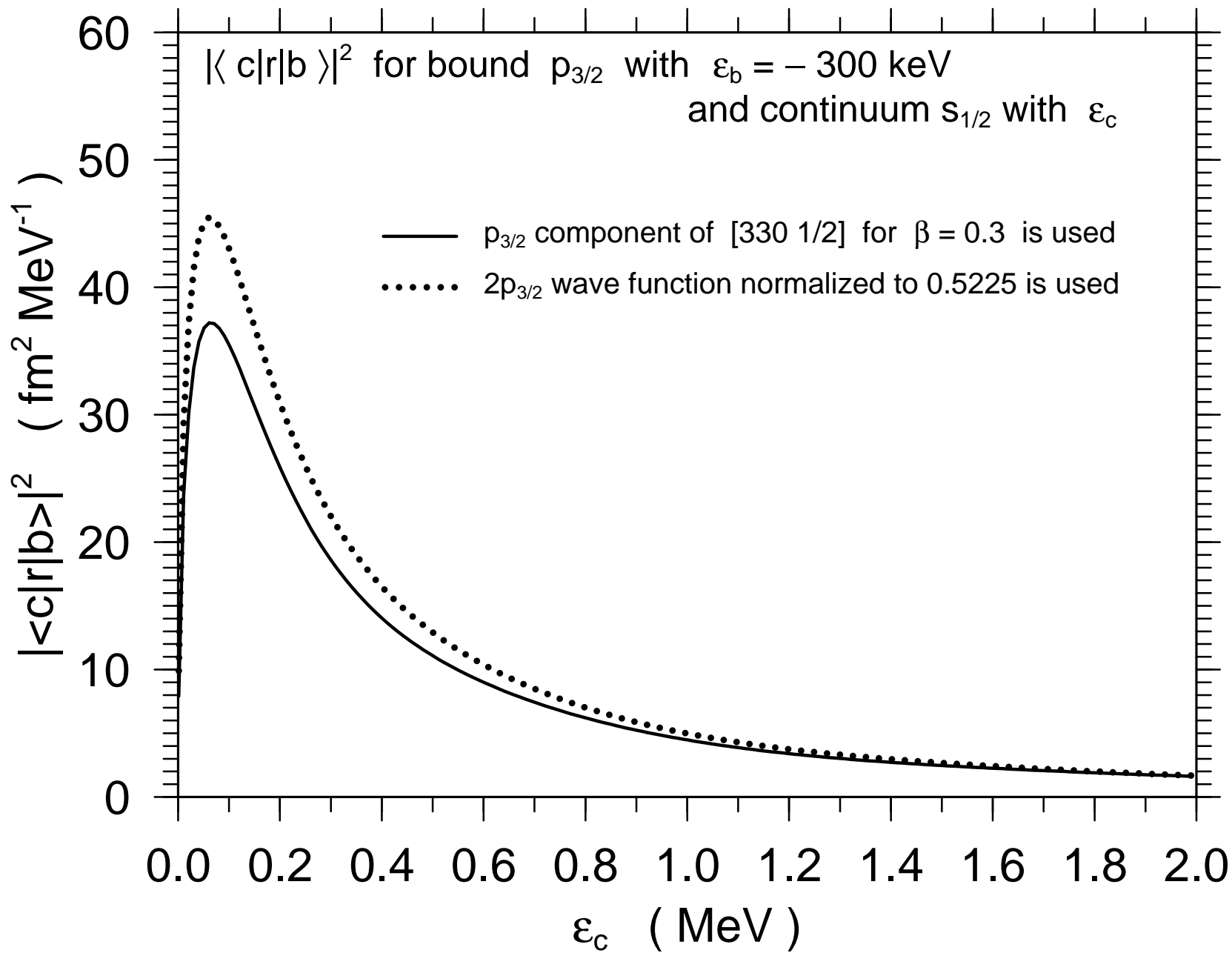
\end{document}